\documentclass[useAMS,usenatbib,onecolumn]{mn2e}
\usepackage{times}
\usepackage{graphicx}
\usepackage[usenames]{color}
\usepackage{ulem}
\begin{document}

\title[Covariance of One-Dimensional Power Spectrum]
{Covariance of the One-Dimensional Mass Power Spectrum}

\author[Zhan \& Eisenstein]
{Hu Zhan$^1$\thanks{Current address: Department of Physics, 
University of California, One Shields Ave, Davis, CA 95616, USA, 
zhan@physics.ucdavis.edu} and 
Daniel Eisenstein$^2$\thanks{deisenstein@as.arizona.edu} \\
$^1$Department of Physics, University of Arizona, Tucson, AZ 85721, USA \\
$^2$Steward Observatory, University of Arizona, Tucson, AZ 85721, USA}

\maketitle

\begin{abstract}

We analyse the covariance of the one-dimensional mass power
spectrum along lines of sight. 
The covariance reveals the correlation between different modes
of fluctuations in the cosmic density field and gives the sample variance
error for measurements of the mass power spectrum. 
For Gaussian random fields, the covariance matrix is diagonal. 
As expected, the variance of the
measured one-dimensional mass power spectrum is inversely proportional to
the number of lines of sight that are sampled from each random field. 
The correlation between lines of sight in a single field may alter
the covariance. However, lines of sight that are sampled far apart 
are only weakly correlated, so that they can be treated as independent
samples. Using $N$-body simulations, we find that
the covariance matrix of the one-dimensional mass power spectrum is
not diagonal for the cosmic density field due to the non-Gaussianity
and that the variance is much higher than that of Gaussian random fields. 
From the covariance, one will be able to
determine the cosmic variance in the measured one-dimensional mass power
spectrum as well as to estimate how many lines of sight are needed to 
achieve a target precision.
\end{abstract}


\section{Introduction}

The one-dimensional mass power spectrum (PS) and its relation to the
three-dimensional mass PS have been frequently utilized to recover the
linear mass PS from the Ly$\alpha$ forest \citep{cwk98, cwp99, cwb02,
gh02, kvh04}. This opens a great window for studying the large-scale
structure of the universe over a wide range of redshift. As one starts to
attempt precision cosmology using the Ly$\alpha$ forest \citep{cwb02,
mms03, svp03}, it becomes necessary to quantify systematic 
uncertainties of the PS analysis including the covariance.

For an ensemble of isotropic fields, the one-dimensional mass PS 
$P_{\rm 1D}(k)$ is a simple integral of the three-dimensional mass PS 
$P_{\rm 3D}(k)$ \citep{lhp89},
\begin{equation} \label{eq:iso321}
P_{\rm 1D}(k)=\frac{1}{2\pi}\int_k^{\infty} P_{\rm 3D}(k^{\prime})
k^{\prime}\,{\rm d}k^{\prime},
\end{equation}
where $k$ is the line-of-sight wavenumber. However, there is only one
observable universe. One has to replace the ensemble average with a
spatial average. For instance, one may sample multiple line-of-sight
densities from the three-dimensional cosmic density field and use the
average PS of the one-dimensional densities in place of the
ensemble-average quantity.

The covariance of the spatial-average PS may differ from that of the
ensemble-average PS for at least two reasons. First, lines of sight 
(LOS's) are no longer independent of each other. Correlations between 
close LOS's could increase the covariance of the one-dimensional mass 
PS, although they may also be used to extract cosmological parameters 
from the Ly$\alpha$ forest \citep{hsb99, mm99, vmm02, m03, rpp03}.
Second, the length of each LOS is always much less than the size
of the universe, so that false correlations between different modes
are introduced in the covariance \citep[for the three-dimensional case,
see][]{fkp94}. Because the cosmic density field is highly non-Gaussian on
scales of interest, $N$-body simulations are helpful for quantifying the
covariance.

The rest of the paper is organized as follows. Section \ref{sec:p3D}
briefly describes the notation and the convention of Fourier
transforms, the three-dimensional mass PS, and its covariance.
The spatial-average one-dimensional mass PS and its covariance are derived
in Sections \ref{sec:321} and \ref{sec:cov}, respectively. The effects of
the line-of-sight length is discussed in Section \ref{sec:short}. Section
\ref{sec:test} presents numerical results of the covariance from $N$-body
simulations with different box sizes. The conclusions and discussions are
given in Section \ref{sec:con}. Unless specified otherwise, the PS refers
to the mass PS.

\section{Three-dimensional Power Spectrum} \label{sec:p3D}

We assume the following convention of Fourier transforms for a cubic
density field of volume $V = B^3$:
\begin{equation} \label{eq:Fourier}
\hat{\delta}(\bmath{n}) = \int_V \delta(\bmath{x}) \,
e^{-2\pi i\bmath{n}\cdot\bmath{x}/B}\,{\rm d}\bmath{x}, \qquad
\delta(\bmath{x}) = \frac{1}{V} \sum_{\bmath{n}=-\infty}^{\infty}
 \hat{\delta}(\bmath{n}) \, e^{2\pi i\bmath{n}\cdot\bmath{x}/B},
\end{equation}
where $\delta(\bmath{x})$ and $\hat{\delta}(\bmath{n})$ are the overdensity
and its Fourier counterpart, respectively, the summation 
$\sum_{\bmath{n}=-\infty}^{\infty}$ is an abbreviation for
$\sum_{n_1,n_2,n_3=-\infty}^{\infty}$, and the wavevector $\bmath{k}=2\pi
\bmath{n}/B$. With the understanding that Fourier modes exist only at
discrete wavenumbers, i.e. $n_1$, $n_2$, and $n_3$ are integers, 
one may use $\bmath{k}$ and $\bmath{n}$ interchangeably
for convenience. To be complete, the orthonormality relations are
\begin{equation} \label{eq:orth_n}
\frac{1}{V} \int_V e^{-2\pi i (\bmath{n}-\bmath{n}')\cdot \bmath{x}/B} 
{\rm d} \bmath{x} = \delta^{\rm K}_{\bmath{n},\bmath{n}'}, \qquad
\label{eq:orth_x} \frac{1}{V} \sum_{\bmath{n}=-\infty}^{\infty} 
e^{2\pi i \bmath{n} \cdot (\bmath{x} - \bmath{x}')/B} 
= \delta^{\rm D}(\bmath{x}-\bmath{x}'), 
\end{equation}
where $\delta^{\rm K}_{\bmath{n},\bmath{n}'}$ is the three-dimensional
Kronecker delta function, and $\delta^{\rm D}(\bmath{x}-\bmath{x}')$ the 
Dirac delta function.

The three-dimensional PS of the universe is defined through
\begin{equation} \label{eq:powsp_n}
\langle \hat{\delta}(\bmath{k})\hat{\delta}^{*}(\bmath{k}') \rangle
= P_{\rm 3D}(\bmath{k}) V \delta^{\rm K}_{\bmath{n},\bmath{n}'},
\end{equation}
where $\langle \ldots \rangle$ stands for an ensemble average. One may
define an observed PS, $\mathcal{P}_{\rm 3D}(\bmath{k}) = 
|\hat{\delta}(\bmath{k})|^2 / V$, so that 
$P_{\rm 3D}(\bmath{k}) = \langle \mathcal{P}_{\rm 3D}(\bmath{k}) \rangle$.
Note that a shot-noise term should be included if the PS is measured 
from discrete objects, and it is 
inversely proportional to the mean number density of the objects. 
We neglect the shot noise in this paper. For an 
isotropic universe, $P_{\rm 3D}(\bmath{k})$ is a function of the length of 
$\bmath{k}$ only, i.e. $P_{\rm 3D}(\bmath{k}) \equiv P_{\rm 3D}(k)$. 
The four-point function of $\hat{\delta}(\bmath{k})$ is
\begin{equation} \label{eq:4point}
\langle \hat{\delta}(\bmath{k})\hat{\delta}^{*}(\bmath{k}') 
\hat{\delta}(\bmath{k}'')\hat{\delta}^{*}(\bmath{k}''')\rangle = V^2
\big [P_{\rm 3D}(k)P_{\rm 3D}(k'')\delta^{\rm K}_{\bmath{n}, \bmath{n}'}
\delta^{\rm K}_{\bmath{n}'', \bmath{n}'''} + 
P_{\rm 3D}(k)P_{\rm 3D}(k')\delta^{\rm K}_{\bmath{n}, \bmath{n}'''}
\delta^{\rm K}_{\bmath{n}', \bmath{n}''} \big] + 
VT(\bmath{k}, -\bmath{k}', \bmath{k}'', -\bmath{k}'''),
\end{equation}
where $T$ is the trispectrum, and we have restricted the wavevectors 
to be in the same hemisphere so that the term 
$P_{\rm 3D}(k)P_{\rm 3D}(k')\delta^{\rm K}_{\bmath{n}, -\bmath{n}''}
\delta^{\rm K}_{\bmath{n}', -\bmath{n}'''}$ does not appear. 
There is a redundancy in the variables of the trispectrum, which has only
six degrees of freedom arising from relative coordinates of the four 
points under the constraint of homogeneity \citep{p80}. 
It is evident from the four-point function that the covariance of the
three-dimensional PS is \citep[see also][]{mw99, ch01}
\begin{equation}
\sigma^2(\bmath{k},\bmath{k}') = \langle 
[\mathcal{P}_{\rm 3D}(\bmath{k})-P_{\rm 3D}(\bmath{k})] [\mathcal{P}_{\rm 3D}(\bmath{k}')-P_{\rm 3D}(\bmath{k}')]\rangle = 
P^2_{\rm 3D}(\bmath{k}) \delta^{\rm K}_{\bmath{n},\bmath{n}'} + 
V^{-1}T(\bmath{k}, -\bmath{k}, \bmath{k}', -\bmath{k}').
\end{equation}
For Gaussian random fields (GRFs), the trispectrum vanishes, and 
$\sigma^2(\bmath{k},\bmath{k}') = 
P^2_{\rm 3D}(\bmath{k}) \delta^{\rm K}_{\bmath{n},\bmath{n}'}$.
In reality, the survey volume is always smaller than the 
observable universe, and the survey geometry is more complex than the
simple case we have assumed. These lead to a modification of the
covariance for GRFs \citep{fkp94}
\begin{equation}
\sigma^2(\bmath{k},\bmath{k}') \simeq P^2_{\rm 3D}(\bmath{k})
\left | \frac{\int u^2(\bmath{x})\, e^{-i (\bmath{k} - \bmath{k}')\, 
\bmath{x}}\, {\rm d}\bmath{x}}{\int u^2(\bmath{x})\,
{\rm d}\bmath{x}} \right|^2,
\end{equation}
where $u(\bmath{x})$ is a weight function that depends on the survey
volume and geometry, and the shot noise is neglected. For the low-redshift 
cosmic density field, the 
non-Gaussianity can dramatically boost the elements of the 
covariance matrix through the trispectrum.

\section{One-dimensional Power Spectrum} \label{sec:321}

For a line-of-sight density that is along the $x_3$-axis and sampled at
$(x_1,x_2)$, the one-dimensional Fourier transform gives
\begin{equation} \label{eq:Fourier1D}
\hat{\delta}(\bmath{x}_\perp,n_3) = \int_0^B \delta(\bmath{x})\, 
e^{-2\pi i n_3x_3/B}\, {\rm d} x_3 =
\frac{1}{B^2} \sum_{\bmath{n}_\perp=-\infty}^{\infty} 
\hat{\delta}(\bmath{n}_\perp,n_3)\, 
e^{2\pi i \bmath{n}_\perp \cdot \bmath{x}_\perp / B},
\end{equation}
where the subscript $\perp$ signifies the first two components of a
vector, i.e. $\bmath{x}_\perp = (x_1,x_2)$. Similar to the 
three-dimensional
PS, the one-dimensional PS is expected to follow
\begin{equation} \label{eq:ens1d}
\langle \hat{\delta}(\bmath{x}_\perp, n_3)
\hat{\delta}^{*}(\bmath{x}_\perp, n'_3) \rangle
= P_{\rm 1D}(n_3) B \delta^{\rm K}_{n_3, n'_3}.
\end{equation}
Substituting equation (\ref{eq:Fourier1D}) in equation (\ref{eq:ens1d})
and making use of equation (\ref{eq:powsp_n}), one finds the relation
between the one-dimensional PS and the three-dimensional PS,
\begin{equation} \label{eq:321sum}
P_{\rm 1D}(n_3) = \frac{1}{B^2} \sum_{\bmath{n}_\perp=-\infty}^{\infty}
P_{\rm 3D}(\bmath{n}_\perp, n_3),
\end{equation}
which is a discrete analog of equation (\ref{eq:iso321}).

Practically, one measures PS's of LOS densities sampled at 
some locations, e.g.~$\mathcal{P}_{\rm 1D}(\bmath{x}_\perp, k_3) =
|\hat{\delta}(\bmath{x}_\perp, k_3)|^2 / B$. A simple estimator of the
one-dimensional PS may be constructed by a spatial average over many
LOS's, i.e. $\mathcal{P}_{\rm 1D}(k_3) = \langle \mathcal{P}_{\rm 1D}(\bmath{x}_\perp, k_3) 
\rangle_{\bmath{x}_\perp}$, where $\langle \ldots \rangle_{\bmath{x}_\perp}$
runs over all LOS's sampled in a single universe. To assess the
performance of the estimator, two questions need to be addressed: (1) How
does $\mathcal{P}_{\rm 1D}(k_3)$ relate itself to $\mathcal{P}_{\rm 3D}(\bmath{k})$; and (2) What is
the covariance of $\mathcal{P}_{\rm 1D}(k_3)$ with respect to $P_{\rm 1D}(k_3)$?
The rest of this section answers the former, and Section \ref{sec:cov} the
latter.

For simplicity, we assume that LOS's are sampled regularly in transverse
directions at an interval of $b = B / m$, where $m$ is an integer. Each
LOS has a length of $B$, and $\bmath{x}_\perp = \bmath{l}_\perp b$ with $l_1,
l_2 = 0, \ldots, m-1$. The estimated one-dimensional PS is then
\begin{equation} \label{eq:raw-P1D}
\mathcal{P}_{\rm 1D}(n_3) = \frac{1}{m^2 B^5} \sum_{\bmath{l}_\perp=0}^{m-1} 
\Big |\sum_{\bmath{n}_\perp = -\infty}^{\infty} 
\hat{\delta}(\bmath{n}_\perp,n_3)\, 
e^{2\pi i \bmath{n}_\perp \cdot \bmath{l}_\perp/m} \Big |^2
= \frac{1}{B^5} \sum_{\bmath{n}_\perp,\, \bmath{j}_\perp=-\infty}^{\infty}
\hat{\delta}(\bmath{n}_\perp,n_3)\,
\hat{\delta}^*(\bmath{n}_\perp\! + m\bmath{j}_\perp,n_3),
\end{equation}
where the equality, 
\begin{equation} \label{eq:suml}
\frac{1}{m} \sum_{l=0}^{m-1} e^{2 \pi i (n-n') l / m} = 
\delta^{\rm K}_{n,n'+mj} \quad j=0,\pm 1, \ldots, \pm \infty,
\end{equation}
has been used to obtain the last line in equation (\ref{eq:raw-P1D}). It
is easy to show with equation (\ref{eq:raw-P1D}) that 
\begin{equation} \label{eq:p1d-ens-x}
P_{\rm 1D}(k_3) = \langle \mathcal{P}_{\rm 1D}(k_3) \rangle = 
\langle \mathcal{P}_{\rm 1D}(\bmath{x}_\perp, k_3) \rangle.
\end{equation}
Note that there is no spatial average on the far right side of equation 
(\ref{eq:p1d-ens-x}). 
It seems that only in the limit 
$m\to \infty$ do $\mathcal{P}_{\rm 1D}(k_3)$
and $\mathcal{P}_{\rm 3D}(\bmath{k})$ follow the same relation as 
equations (\ref{eq:iso321}) and (\ref{eq:321sum}), 
i.e.~$\mathcal{P}_{\rm 1D}(k_3) = \sum_{\bmath{n}_\perp=-\infty}^{\infty} 
\mathcal{P}_{\rm 3D}(\bmath{k}_\perp, k_3) / B^2$, 
where we have assumed $\hat{\delta}(\infty) = 0$ so that only 
$\bmath{j}_\perp = (0, 0)$ terms contribute. In other words, to reduce
the uncertainties of the recovered $\mathcal{P}_{\rm 3D}(\bmath{k})$ and 
$P_{\rm 3D}(\bmath{k})$, one has to increase the sampling rate in 
the transverse direction. 

Without losing generality, one may choose $m$ to be even, so that equation
(\ref{eq:raw-P1D}) can be re-arranged into
\begin{equation} \label{eq:est-P1D}
\mathcal{P}_{\rm 1D}(n_3) = \frac{1}{B^2} \sum_{\bmath{n}_\perp=-m/2}
^{m/2} \mathcal{P}^{\rm a}_{\rm 3D}(\bmath{n}_\perp, n_3) =
\frac{1}{B^2} \sum_{\bmath{n}_\perp=-\infty}^{\infty} 
\mathcal{P}_{\rm 3D}(\bmath{n}_\perp, n_3) + A(n_3),
\end{equation}
where
\begin{equation} \label{eq:aliasPS}
\mathcal{P}^{\rm a}_{\rm 3D}(\bmath{n}_\perp, n_3) = \frac{1}{V} 
\Big|\sum_{\bmath{j}_\perp=-\infty}^{\infty}
\hat{\delta}(\bmath{n}_\perp\! +m\bmath{j}_\perp,n_3)\Big|^2,
\end{equation}
and
\begin{equation} \label{eq:A}
A(n_3) = \frac{1}{B^5} \sum_{\mbox{\scriptsize $\begin{array}{c}
\bmath{j}_\perp,\bmath{j}'_\perp\! = -\infty \\ 
\bmath{j}'_\perp \!\ne 0 \end{array}$}}^{\infty}
\sum_{\bmath{n}_\perp\!=-m/2}^{m/2} 
\hat{\delta}(\bmath{n}_\perp\! + m\bmath{j}_\perp,n_3) \,
\hat{\delta}^*[\bmath{n}_\perp \! + m(\bmath{j}_\perp
\! + \bmath{j}'_\perp),n_3].
\end{equation}
Equation (\ref{eq:est-P1D}) provides two equivalent views of the sampling
effect. On one hand, the one-dimensional PS is a sum of the aliased
three-dimensional PS, $\mathcal{P}^{\rm a}_{\rm 3D}(\bmath{k})$, that is 
sampled by a grid of 
$m\times m$ LOS's. On the other hand, it is a complete sum of the underlying
three-dimensional PS with an extra term $A(k_3)$ that is determined by the
sampling rate and properties of the density field.

The discrete Fourier transform cannot distinguish a principal mode at 
$|k| \le k_{\rm Nyq}$ from its aliases at $k\pm 2k_{\rm Nyq},
k\pm 4k_{\rm Nyq},\ldots$, where $k_{\rm Nyq}$ is the sampling Nyquist 
wavenumber. For example, $k_{\rm Nyq} = \pi / b$ if one samples the 
field at an equal spacing of $b$. The alias modes will be added to the 
principal mode if they are not properly filtered out before sampling
\citep[for details, see][]{he81}. Since the cosmic density field is not
band-limited, aliasing can distort statistics of the field. The 
distortion on the three-dimensional PS cannot be quantified \emph{a priori}, 
because it depends on relative phases between the principal mode 
and its aliases. The alias effect is less pronounced, if amplitudes of 
the alias modes are much smaller than that of the principal mode. Since 
the three-dimensional PS decreases toward small scales, a high 
sampling rate (or $k_{\rm Nyq}$) can suppress aliasing for modes with 
$k \ll k_{\rm Nyq}$. 
%Alternatively, one may
%use anti-aliasing filters to reduce the alias effect. 

%Aliasing occurs in the estimated one-dimensional PS because the 
%continuous 
%density field contains significant Fourier modes at wavenumbers that 
%are greater than the sampling Nyquist wavenumber in $x_1$ and $x_2$ 
%directions. This is evident in equation (\ref{eq:aliasPS}). Unlike the
%three-dimensional case, prefiltering small-scale transverse modes may 
%not improve the estimated one-dimensional PS, because each mode of the 
%theoretical one-dimensional PS contains contributions of transverse 
%modes in the three-dimensional cosmic density field on all scales. 

If the Fourier modes of the density field are uncorrelated, the term 
$A(k_3)$ may be neglected even for a finite number of LOS's and, therefore, 
validate equation (\ref{eq:321sum}). Strictly speaking, $A(k_3)$ 
vanishes only as an ensemble average over many GRFs, but since there 
are so many independent modes in a shell of radius around 
$k$ -- especially at large wavenumbers -- the summation in $A(k_3)$ 
will tend to vanish even for a single GRF. The cosmic density field is
more Gaussian at higher redshift, so equation (\ref{eq:321sum}) may be a
good approximation then. At low redshift, however, the non-Gaussianity 
alters the statistics of one-dimensional fields \citep[see e.g.][]{a94}, 
such that one might only be able to recover a heavily aliased 
three-dimensional PS from a sparse sample of LOS's.

\section{Covariance of the One-dimensional Power Spectrum} \label{sec:cov}

The covariance of the PS is of interest because it tells us how
likely an estimated PS is to represent the true PS and how much the modes 
on different scales are correlated. For the one-dimensional PS, the
covariance is defined as
\begin{equation} \label{eq:cov-def}
\sigma^2_{\rm 1D}(k_3, k_3') = \langle [\mathcal{P}_{\rm 1D}(k_3) - P_{\rm 1D}(k_3)]
[\mathcal{P}_{\rm 1D}(k_3') - P_{\rm 1D}(k_3')] \rangle.
\end{equation}
The covariance of the mean PS of $N$ LOS's that are sampled in a single 
field can be expanded into a sum of 
pair-wise covariances between two LOS's separated by 
$\bmath{s}^{jl}_\perp = \bmath{x}^j_\perp - \bmath{x}^l_\perp$, e.g.
\begin{equation} \label{eq:cov2}
\sigma^2_{\rm 1D}(k_3, k_3') = \frac{1}{N^2} \sum_{j, l = 1}^{N} 
\sigma^2_{\rm 1D}(k_3, k_3'; \bmath{s}^{jl}_\perp),
\end{equation}
where 
\[
\sigma^2_{\rm 1D}(k_3, k_3'; \bmath{s}^{jl}_\perp)
 = \langle [\mathcal{P}_{\rm 1D}(\bmath{x}^j_\perp, k_3) - P_{\rm 1D}(k_3)]
[\mathcal{P}_{\rm 1D}(\bmath{x}^l_\perp, k_3') - P_{\rm 1D}(k_3')] \rangle ,
\]
and $\bmath{x}^j_\perp$ is the location of the $j$th LOS in the
$x_1$--$x_2$ plane.

For GRFs, the four-point function [equation (\ref{eq:4point})] helps
reduce the pair-wise covariance to
\begin{eqnarray} \label{eq:cov2-xi} \nonumber
\sigma^2_{\rm 1D}(n_3, n_3';\bmath{s}) &=& \frac{1}{B^{10}} 
\!\sum_{\bmath{n}_\perp, \bmath{n}'_\perp, \bmath{n}''_\perp, \bmath{n}'''_\perp 
= -\infty}^{\infty} \!\!\!\langle \hat{\delta}(\bmath{n}_\perp, n_3)
\hat{\delta}^*(\bmath{n}'_\perp, n_3)\hat{\delta}(\bmath{n}''_\perp, n'_3)
\hat{\delta}^*(\bmath{n}'''_\perp, n'_3) \rangle \\ \nonumber
&& \qquad\qquad\quad \times e^{2\pi i [(\bmath{n}_\perp - \bmath{n}'_\perp)
\cdot \bmath{x}_\perp^j + (\bmath{n}''_\perp - \bmath{n}'''_\perp)\cdot 
\bmath{x}_\perp^l]/B} - P_{\rm 1D}(n_3)P_{\rm 1D}(n_3') \\
&=& \Big |\frac{1}{B^2} \sum_{\bmath{n}_\perp = -\infty}^{\infty} 
P_{\rm 3D}(\bmath{n}_\perp, n_3)\, e^{2\pi i \bmath{n}_\perp \cdot\, \bmath{s} / B} 
\Big |^2 \delta^{\rm K}_{n_3, n_3'} \equiv |\xi(\bmath{s}, n_3)|^2
\delta^{\rm K}_{n_3, n_3'},
\end{eqnarray} 
where the subscript and superscript are dropped for $\bmath{s}$. 
Fourier transforms
of $\xi(\bmath{s}, k_3)$ will give the three-dimensional PS and correlation
function of the density field, and $P_{\rm 1D}(k_3) = \xi(0, k_3)$
\citep[see also][]{vmm02}. Because of isotropy, the pair-wise covariance 
$\sigma^2_{\rm 1D}(k_3, k_3';\bmath{s})$ and $\xi(\bmath{s}, k_3)$ depend 
only on the magnitude of the separation, 
i.e.~$\sigma^2_{\rm 1D}(k_3, k_3';\bmath{s}) \equiv \sigma^2_{\rm
1D}(k_3, k_3';s)$ and $\xi(\bmath{s}, k_3) \equiv \xi(s, k_3)$.

If one LOS is sampled from each GRF, the variance of the measured
one-dimensional PS $\mathcal{P}_{\rm 1D}(k_3)$ is $\sigma^2_{\rm 1D}(k_3, k_3) =
\sigma^2_{\rm 1D}(k_3, k_3; 0) = P^2_{\rm 1D}(k_3)$, analogous to the
three-dimensional case. If $N=m^2$ LOS's are sampled in each GRF on a
regular grid as in Section \ref{sec:321}, e.g.~$\bmath{s}$ = $(\bmath{l}_\perp
- \bmath{l}'_\perp) b$, the covariance becomes
\begin{eqnarray} \label{eq:cov-sum}
\sigma^2_{\rm 1D}(n_3, n_3') &=& \frac{1}{N^2}
\sum_{\bmath{l}_\perp, \bmath{l}'_\perp=0}^{m-1} 
\sigma^2_{\rm 1D}[n_3, n_3'; (\bmath{l}_\perp - \bmath{l}'_\perp)b] 
= \frac{1}{N^2 B^4} 
\sum_{\bmath{n}_\perp, \bmath{n}'_\perp = -\infty}^{\infty} \!\!\!\!\! 
P_{\rm 3D}(\bmath{n}_\perp, n_3) P_{\rm 3D}(\bmath{n}'_\perp, n_3) \,
\Big | \sum_{\bmath{l}_\perp=0}^{m-1} 
e^{2\pi i (\bmath{n}_\perp - \bmath{n}'_\perp) \cdot \bmath{l}_\perp / m} 
\Big |^2 \delta^{\rm K}_{n_3, n_3'} \nonumber \\ &=& \frac{1}{B^4} 
\sum_{\bmath{n}_\perp,\, \bmath{j}_\perp = -\infty}^{\infty} \!\!\!\!\! 
P_{\rm 3D}(\bmath{n}_\perp, n_3) 
P_{\rm 3D}(\bmath{n}_\perp+m\bmath{j}_\perp, n_3) 
\,\delta^{\rm K}_{n_3, n_3'},
\end{eqnarray}
where we have used equation (\ref{eq:suml}) to reach the last line. This
result can be easily obtained using equations (\ref{eq:4point}) and 
(\ref{eq:raw-P1D}) as well.
It is seen that the summation in the last line of equation
(\ref{eq:cov-sum}) runs over only one mode out of every $N$ modes in the
Fourier space. If the three-dimensional PS were constant, the variance of
the mean PS of the $N$ LOS's would be $N$ times smaller than the variance
of the PS of a single LOS. This is coincident
with the theory of the variance of the mean. Since the cosmic density
field is not a GRF in general, equation (\ref{eq:cov-sum}) is not expected
to give an accurate estimate.

If the $N$ LOS's are sampled randomly in each GRF, the summation over
LOS's in the second line of equation (\ref{eq:cov-sum}) should be
re-cast to read
\begin{equation} \label{eq:cov-ran}
\sigma^2_{\rm 1D}(n_3, n_3') = \frac{1}{N^2 B^4} 
\sum_{\bmath{n}_\perp, \bmath{n}'_\perp = -\infty}^{\infty}
\!\!\!\!\! P_{\rm 3D}(\bmath{n}_\perp, n_3) P_{\rm 3D}(\bmath{n}'_\perp, n_3)
\sum_{j,\, l=1}^{N} e^{2\pi i (\bmath{n}_\perp - \bmath{n}'_\perp) \cdot 
(\bmath{x}^j_\perp - \bmath{x}^l_\perp)/B} \,\delta^{\rm K}_{n_3, n_3'}.
\end{equation}
Since $\bmath{x}^j_\perp$ is randomly distributed, the second sum in 
equation (\ref{eq:cov-ran}) tends to vanish for a large number of 
LOS's except when $j=l$. Thus, one obtains
\begin{equation} \label{eq:cov-ran2}
\sigma^2_{\rm 1D}(n_3, n_3') \simeq \frac{1}{N B^4} 
\sum_{\bmath{n}_\perp, \bmath{n}'_\perp = -\infty}^{\infty}
\!\!\!\!\! P_{\rm 3D}(\bmath{n}_\perp, n_3) P_{\rm 3D}(\bmath{n}'_\perp, n_3)\,
\delta^{\rm K}_{n_3, n_3'} = \frac{1}{N} P^2_{\rm 1D}(n_3) \,
\delta^{\rm K}_{n_3, n_3'}.
\end{equation}
Again, it shows that the variance of $\mathcal{P}_{\rm 1D}(k_3)$ is inversely
proportional to the number of LOS's, but this is valid only for a large
number of LOS's randomly sampled in a GRF.

Through the Gaussian case one can find the lowest bound of the 
uncertainty for a measured one-dimensional PS and the minimum number of 
LOS's needed for a target precision. For example, to measure the 
one-dimensional PS of GRFs accurate to $5\%$ on \emph{every} mode, one 
needs at least 400 
LOS's. However, the cosmic density field has a far more complex 
covariance due to its non-vanishing trispectrum, and it can have a much
larger variance of the measured one-dimensional PS. 

The trispectrum introduces an extra term to the covariance of the 
one-dimensional 
 PS in addition to the Gaussian piece [equation (\ref{eq:cov-sum})], i.e.
\begin{equation} \label{eq:cov-tri}
\sigma^2_{\rm 1D}(n_3, n_3') = \frac{1}{B^4} 
\sum_{\bmath{n}_\perp,\, \bmath{j}_\perp = -\infty}^{\infty}
\!\!\!\!\! P_{\rm 3D}(\bmath{n}_\perp, n_3) P_{\rm 3D}(\bmath{n}_\perp+m\bmath{j}_\perp, n_3) 
\,\delta^{\rm K}_{n_3, n_3'} + \frac{1}{B}T_{\rm 1D}(n_3, n_3'),
\end{equation}
where 
\begin{equation} \label{eq:T1D}
T_{\rm 1D}(n_3, n_3') = \frac{1}{B^6} 
\sum_{\bmath{n}_\perp, \,\bmath{j}_\perp,\, \bmath{n}'_\perp, 
\,\bmath{j}'_\perp = -\infty}^{\infty} 
T(\bmath{n}, -\bmath{n} - m\bmath{j}, \bmath{n}', -\bmath{n}' - m\bmath{j}'),
\end{equation}
$\bmath{n}_\perp$, $\bmath{j}_\perp$, $\bmath{n}'_\perp$, and 
$\bmath{j}'_\perp$ are the transverse components of 
$\bmath{n}$, $\bmath{j}$, $\bmath{n}'$, and 
$\bmath{j}'$, respectively, and $j_3 = j'_3 = 0$. 
In deriving equation (\ref{eq:cov-tri}), we have made use of
equations (\ref{eq:4point}) and (\ref{eq:raw-P1D}).

\section{Line-of-Sight Length} \label{sec:short}

Observationally, the length of a LOS is always much less than the size of
the observable universe. For example, in the case of a pencil-beam 
survey, the length is limited by the depth of the survey. For the 
Ly$\alpha$ forest, it is often determined by the distance between the 
starting redshift of Ly$\alpha$ and Ly$\beta$ lines. Damped Ly$\alpha$ 
systems and other astrophysical or instrumental factors can break the 
spectrum into yet shorter chunks. Even if a very long LOS were 
obtained, one might still wish to measure the PS on smaller segments to
avoid evolutionary effects. To account for the length of LOS's, the 
one-dimensional PS and its covariance must be re-formulated.

For a LOS from $(\bmath{x}_\perp, 0)$ to $(\bmath{x}_\perp, L)$, its 
Fourier transform is
\begin{equation} 
\tilde{\delta}(\bmath{x}_\perp, \tilde{n}_3) 
= \int_0^L \delta (\bmath{x}_\perp, x_3) 
e^{-2\pi i \tilde{n}_3 x_3 / L} {\rm d} x_3 
= \frac{L}{B^3} \sum_{\bmath{n} = -\infty}^{\infty} 
\hat{\delta}(\bmath{n}) e^{2\pi i \bmath{n}_\perp \cdot
\bmath{x}_\perp / B} 
\frac{e^{2\pi i (n_3 L / B - \tilde{n}_3)} - 1} 
{2\pi i (n_3 L / B - \tilde{n}_3)},
\end{equation}
so that
\begin{equation}
\langle \tilde{\delta}(\bmath{x}_\perp, \tilde{n}_3)
\tilde{\delta}^*(\bmath{x}_\perp, \tilde{n}'_3) \rangle = 
\frac{L^2}{B^3} \sum_{\bmath{n} = -\infty}^{\infty} 
P_{\rm 3D}(\bmath{n}) w(n_3, \tilde{n}_3) w(n_3, \tilde{n}'_3),
\end{equation}
where 
\begin{equation} \label{eq:windlen}
w(n_3, \tilde{n}_3) = (-1)^{\tilde{n}_3} \frac{\sin[(n_3L/B-\tilde{n}_3)\pi]}
{(n_3L/B-\tilde{n}_3)\pi }.
\end{equation}
The function $w(n_3, \tilde{n}_3)$ is a window function that mixes
Fourier modes of the density field along $n_3$ direction into the
one-dimensional Fourier mode at $\tilde{n}_3$. Because $w(n_3,
\tilde{n}_3)$ and $w(n_3, \tilde{n}'_3)$ are not orthogonal to each other,
the term $\langle \tilde{\delta}(\bmath{x}_\perp, \tilde{n}_3)
\tilde{\delta}^*(\bmath{x}_\perp, \tilde{n}'_3) \rangle$ is no
longer diagonal with respect to $\tilde{n}_3$ and $\tilde{n}'_3$. However,
it remains diagonal-dominant.

One may define the observed one-dimensional PS at $\bmath{x}_\perp$ as
\begin{equation}
\widetilde{\mathcal{P}}_{\rm 1D} (\bmath{x}_\perp, \tilde{n}_3) = \langle
|\tilde{\delta}(\bmath{x}_\perp, \tilde{n}_3)|^2 \rangle / L. 
\end{equation}
The ensemble-averaged one-dimensional PS is
\begin{equation}
\widetilde{P}_{\rm 1D}(\tilde{n}_3) = \langle 
\widetilde{\mathcal{P}}_{\rm 1D}(\bmath{x}_\perp, \tilde{n}_3) \rangle =
\frac{L}{B^3} \sum_{\bmath{n} = -\infty}^{\infty} 
P_{\rm 3D}(\bmath{n}) w^2(n_3, \tilde{n}_3).
\end{equation}
For GRFs, the covariance of 
$\widetilde{\mathcal{P}}_{\rm 1D} (\tilde{n}_3)$ is
\begin{eqnarray} \label{eq:cov-short}
\tilde{\sigma}^2_{\rm 1D}(\tilde{n}_3, \tilde{n}'_3) &= &
\langle [\widetilde{\mathcal{P}}_{\rm 1D}(\bmath{x}_\perp, \tilde{n}_3)
- \widetilde{P}_{\rm 1D}(\tilde{n}_3)]
[\widetilde{\mathcal{P}}_{\rm 1D}(\bmath{x}_\perp, \tilde{n}'_3) -
\widetilde{P}_{\rm 1D}(\tilde{n}'_3)] \rangle \nonumber \\ &=&
\frac{L^2}{B^6} \!\sum_{\bmath{n},\bmath{n}' = -\infty}^{\infty} \!\!
P_{\rm 3D}(\bmath{n})P_{\rm 3D}(\bmath{n}') w(n_3, \tilde{n}_3) w(n_3, \tilde{n}'_3)
 w(n'_3, \tilde{n}_3) w(n'_3, \tilde{n}'_3).
\end{eqnarray}
As expected, the covariance is not diagonal because of the window function
$w(n_3, \tilde{n}_3)$.

\section{$N$-body Tests} \label{sec:test}
Even at moderately high redshift, the cosmic density field is already 
quite non-Gaussian on scales below 10 \mbox{$h^{-1}$Mpc}.
When projected in one dimension, the small-scale non-Gaussianity will 
obviously affect the measured one-dimensional PS on much larger scales. 
The non-vanishing trispectrum introduces an extra term, 
$T_{\rm 1D}(k_3, k'_3)$, to the covariance of the one-dimensional PS
[see equation (\ref{eq:cov-tri})]. Although it is possible to 
derive an approximate trispectrum based on the halo model 
\citep[e.g.][]{ch01}, the one-dimensional projection, unfortunately, obscures 
the contribution of the trispectrum to the one-dimensional PS. 
Therefore, numerical simulations are necessary for the study.

Three $N$-body simulations of 256$^3$ cold dark matter (CDM) particles are
used to quantify the covariance of the one-dimensional PS. The model
parameters are largely consistent with \emph{WMAP} results \citep{svp03},
e.g.~($\Omega$, $\Omega_{\rm b}$, $\Omega_\Lambda$, $h$, $\sigma_8$, $n$) 
= (0.27, 0.04, 0.73, 0.71, 0.85, 1), where $\Omega$ is the cosmic matter
density parameter, $\Omega_{\rm b}$ the baryon density parameter, 
$\Omega_\Lambda$ the energy density parameter associated with the 
cosmological constant, $\sigma_8$ the rms density
fluctuation within a radius of 8 \mbox{$h^{-1}$Mpc}, and $n$ is the power
spectral index. The box sizes of the simulations are 128
\mbox{$h^{-1}$Mpc} (labelled as B128), 256 \mbox{$h^{-1}$Mpc} (B256), and
512 \mbox{$h^{-1}$Mpc} (B512). The baryon density parameter 
$\Omega_{\rm b}$ is
used only for the purpose of calculating the transfer function using
\textsc{linger} \citep{mb95}, which is then read by \textsc{grafic2}
\citep{b01} to generate the initial condition. The CDM particles are
evolved from $z = 44.5$ to the present using \textsc{gadget} 
\citep{syw01}. Additionally, GRFs are generated with the same 
box sizes as those of $N$-body simulations for comparison and are 
labelled as R128, R256, and R512, respectively.

The simulations produce snapshots at $z=3$ and 0. For each snapshot, the
particles are assigned to a density grid of 512$^3$ nodes using the
triangular-shaped-cloud scheme \citep{he81}. LOS's are then sampled
on the $x_1$--$x_2$ plane, and each LOS is extracted with full resolution
along the $x_3$-axis. 
Because the density grid has a finite resolution, both the coordinates 
and the wavenumbers are discrete and finite. The equations in Sections
\ref{sec:p3D}--\ref{sec:short} need to be modified accordingly, so that
they do not sum over non-existing modes. For example, equation
(\ref{eq:321sum}) becomes
\begin{equation} \label{eq:321fsum}
P_{\rm 1D}(n_3) = \frac{1}{B^2} 
\sum_{\bmath{n}_\perp=-M/2}^{M/2}P_{\rm 3D}(\bmath{n}_\perp,n_3),
\end{equation}
where $M = 512$ is the number of nodes of the density grid in transverse
directions.

\begin{figure}
\centering
\includegraphics[width=6in]{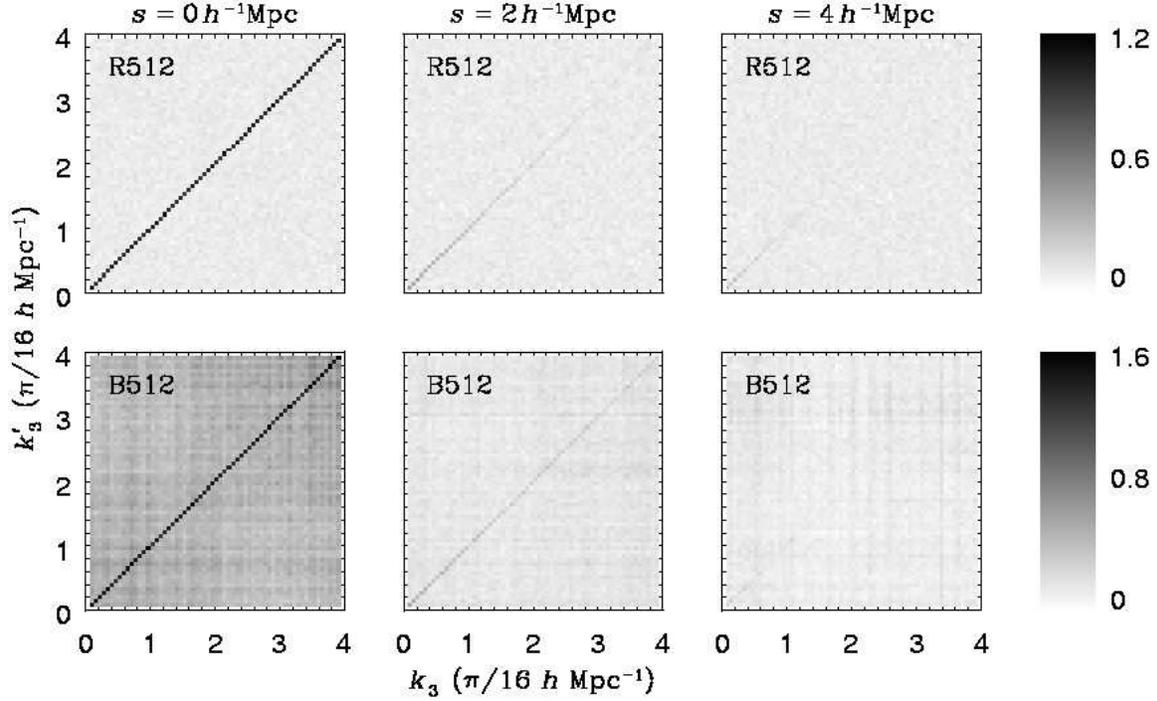}
\caption{Normalized pair-wise covariance matrices $C(k_3, k_3'; s)$ 
[see equation (\ref{eq:Cs})] in grey scale. 
The covariances are calculated over 2000 pairs of LOS's sampled at a 
fixed separation $s$ from the simulation B512 at $z=3$ (lower panels)
and from 2000 GRFs (R512) that have the same box size 
and three-dimensional mass PS as the simulation (upper panels).
The grey scale extends to $-0.1$, which corresponds to white.
\label{fig:cov2}}
\end{figure}

\begin{figure}
\centering
\includegraphics[width=5in]{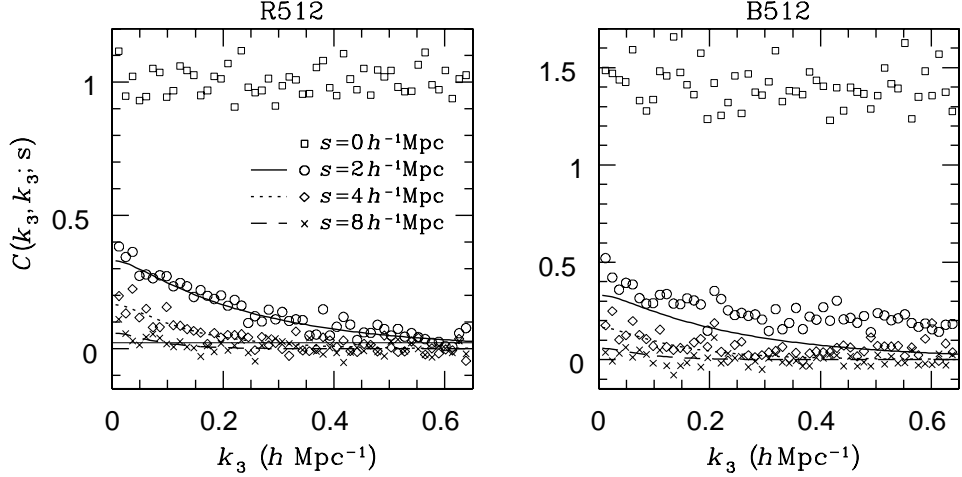}
\caption{Diagonal elements of the covariance matrices $C(k_3, k_3; s)$. 
The horizontal
solid line in the left panel marks the rms value of the off-diagonal
elements in the $s = 8\ h^{-1}$Mpc case. Other lines are from direct
summations of the three-dimensional mass PS using equation 
(\ref{eq:cov2-xi}).
The symbols are measured from GRFs R512 (left) and the 
simulation B512 at $z=3$
(right). For GRFs, $C(k_3, k_3; 0)$ is expected to be unity.
\label{fig:xi}}
\end{figure}

For brevity and the purpose of comparing the covariance, we
introduce the following three covariance matrices: 
(1) The normalized pair-wise covariance between two LOS's,
\begin{equation} \label{eq:Cs}
C(k_3, k_3'; s) = \sigma^2_{\rm 1D}(k_3, k_3'; s) 
[P_{\rm 1D}(k_3)P_{\rm 1D}(k_3')]^{-1},
\end{equation}
where $s$ is the transverse separation between the two LOS's.
(2) The normalized covariance of the estimated one-dimensional PS,
\begin{equation} \label{eq:C}
C(k_3, k_3') = \sigma^2_{\rm 1D}(k_3, k_3') 
[P_{\rm 1D}(k_3)P_{\rm 1D}(k_3')]^{-1},
\end{equation}
where $\sigma^2_{\rm 1D}(k_3, k_3')$ is the covariance of the mean 
PS of $N$ LOS's, e.g.~equation (\ref{eq:raw-P1D}), and we have omitted
its dependence on the number of LOS's. (3) The reduced covariance
\begin{equation} \label{eq:Chat}
\hat{C}(k_3, k_3') = C(k_3, k_3') [C(k_3, k_3)C(k_3', k_3')]^{-1/2}. 
\end{equation} 
For GRFs, all these covariances are diagonal in matrix representation, 
if the length of LOS's is the same as the size of the simulation box. In
addition, we have $C(k_3, k_3; 0) = 1$ and $C(k_3, k_3) = N^{-1}$, 
where $N$ is the number of LOS's that are sampled for estimating the 
one-dimensional PS. 
The advantage of $\hat{C}(k_3, k_3')$ is that $\hat{C}(k_3, k_3) =
1$ for all fields, so that they can be compared with each other in a
single (grey) scale.

The normalized pair-wise covariance $C(k_3, k_3'; s)$ is shown in
Fig.~\ref{fig:cov2} for GRFs R512 and the simulation B512 at 
$z = 3$. The covariances for GRFs are averaged over an ensemble of 2000 
random
realizations, while those for the simulation are averaged over 2000 pairs
of LOS's from a single field. The behaviour of the covariances is consistent
with the expectation. Namely, $C(k_3, k_3'; 0) \simeq \delta^{\rm K}_{n_3,
n_3'}$ and $C(k_3, k_3; s)$ decreases as the separation $s$ increases. The
simulation does deviate from GRFs because of the non-Gaussianity, which 
increases the variance $C(k_3, k_3; s)$. Fig.~\ref{fig:xi} compares 
the diagonal elements $C(k_3, k_3; s)$ with those calculated using 
equation (\ref{eq:cov2-xi}). We note in passing that the expected values 
of $C(k_3, k_3;8\ \mbox{$h^{-1}$Mpc})$ are 
so close to 0 that they are even below the rms value of the off-diagonal 
elements of $C(k_3, k'_3;8\ \mbox{$h^{-1}$Mpc})$ for 2000 GRFs.
Hence, it is practically difficult to recover three-dimensional
statistics from $\sigma_{\rm 1D}(k_3, k_3; s)$ or $\xi(s, k_3)$ if the
LOS's are too far apart. In theory, the modes of two LOS's are always 
correlated as $k_3 \to 0$, regardless of their separation. However, 
large-scale LOS modes are greatly affected by small-scale 
three-dimensional perturbations, which are nearly-uncorrelated for 
large separations. As such, the 
correlation for $s \ga 8\ h^{-1}$Mpc is so weak that it will not be 
easily detected against statistical uncertainties even for a sizable
sample of 2000 pairs of LOS's. Hence,
Figs.~\ref{fig:cov2} and \ref{fig:xi} suggest that LOS's sampled in a 
single cosmic density field are practically independent of each other as 
long as they are well separated. This 
condition is easily met. For example, \citet{msb04} use 3035 $z > 2.3$
quasar spectra from the Sloan Digital Sky Survey (SDSS, over 2627 
square degrees) to determine the Ly$\alpha$ flux PS. Assuming 
\emph{WMAP} parameters, one will find that the mean separation between 
LOS's is $\sim 60$ comoving $h^{-1}$Mpc such that the covariance 
arisen from the correlation between LOS's can be safely neglected.

\begin{figure}
\centering
\includegraphics[width=5in]{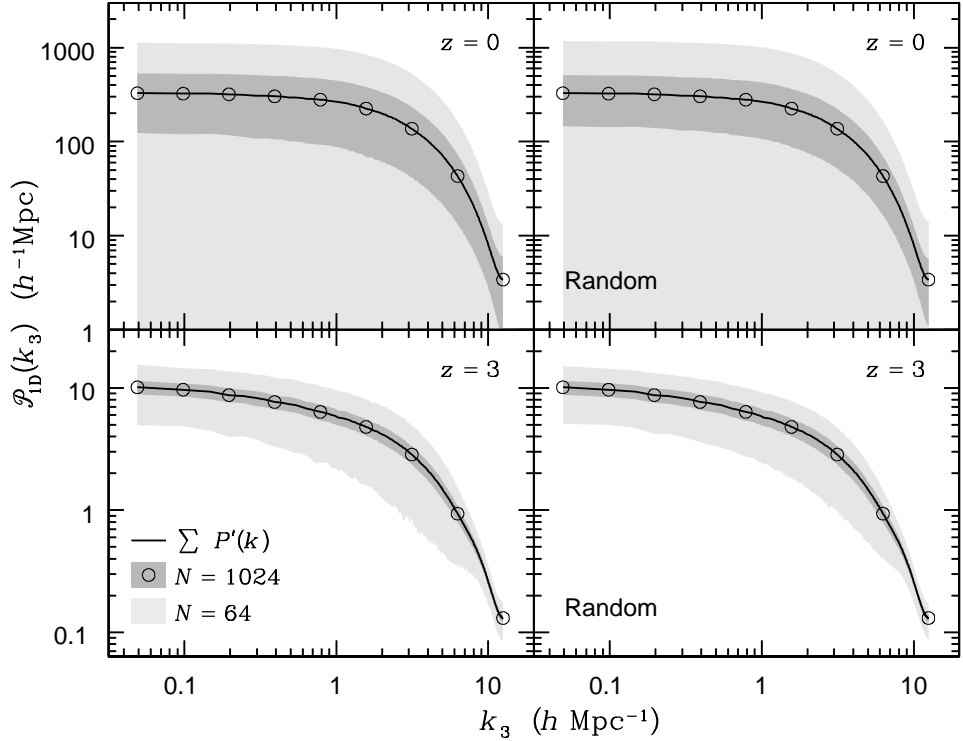}
\caption{Estimated one-dimensional mass PS's $\mathcal{P}_{\rm 1D}(k_3)$. 
The PS's are measured by
averaging over 64 (light grey) and 1024 (dark grey) LOS's that are drawn
from the B128 simulation. Shaded areas mark $1 \sigma$ dispersions of
the PS's among 2000 (light grey) and 256 (dark grey, as $256 \times 1024$
exhausts all the $512^2$ LOS's) distinct drawings. Circles are the
mean PS's for dark grey areas, i.e.~the mean of $512^2$ LOS's. Solid
lines are results of a direct summation of the three-dimensional mass PS using
equation (\ref{eq:321fsum}). The LOS's are sampled on a grid with fixed 
spacing in the left panels but drawn randomly in the right panels.
\label{fig:var1d}}
\end{figure}

\begin{figure} 
\centering
\includegraphics[width=6in]{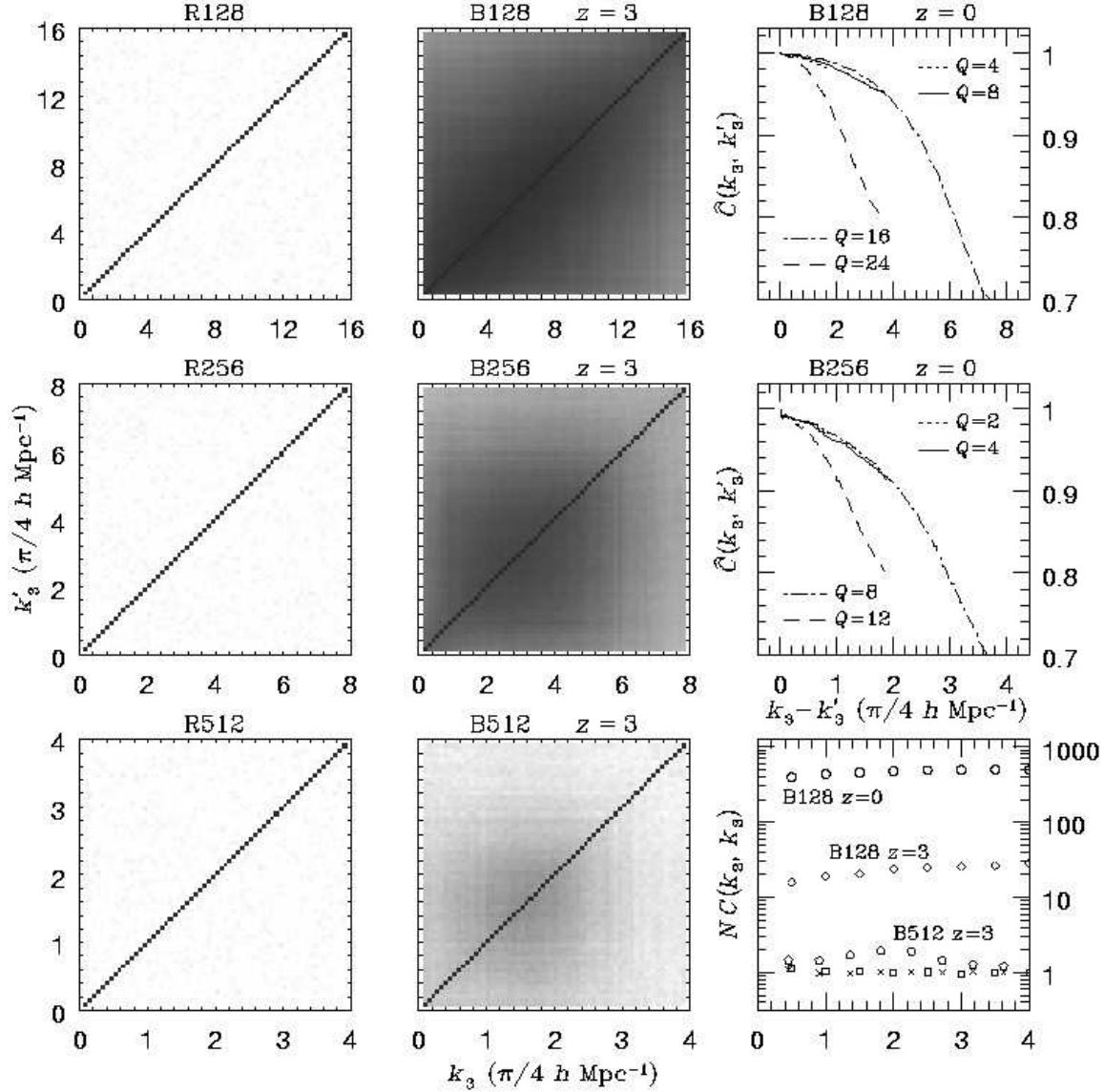}
\caption{Reduced covariances $\hat{C}(k_3,k'_3)$ 
[see equation (\ref{eq:Chat})] averaged over 2000 groups,
each of which consists of 64 LOS's ($N=64$). For each panel in the left,
the LOS's are randomly drawn from a single GRF that has a box size of 128
\mbox{$h^{-1}$Mpc} (R128), or 256 \mbox{$h^{-1}$Mpc} (R256), or 512
\mbox{$h^{-1}$Mpc} (R512). The GRFs have the same three-dimensional mass PS as
their corresponding simulations at $z=0$, but note that
$\hat{C}(k_3,k'_3)$ is independent of redshift for GRFs. Similarly, the
middle column is for simulations at $z=3$. The covariances are shown
in a linear grey scale with black being $1.2$ and white less than or equal
to 0. At $z = 0$, the normalized covariances become much less diagonally
dominant than the B128 $z=3$ panel in the same grey scale, so only 4 cross
sections along $Q=(k_3+k_3')/(\pi/4\ h\ \mbox{Mpc}^{-1})$ are plotted for
B128 and B256. Diagonal elements $C(k_3, k_3)$ 
[see equation (\ref{eq:C})] of R128 (squares), B128
at $z=0$ (circles), B128 at $z=3$ (diamonds), R512 (crosses), and B512 at
$z=3$ (pentagons) are shown with a multiplication factor $N$ in the lower 
right panel. For GRFs, $N C(k_3, k_3) = 1$.
\label{fig:cov64}}
\end{figure}

\begin{figure} 
\centering
\includegraphics[width=6in]{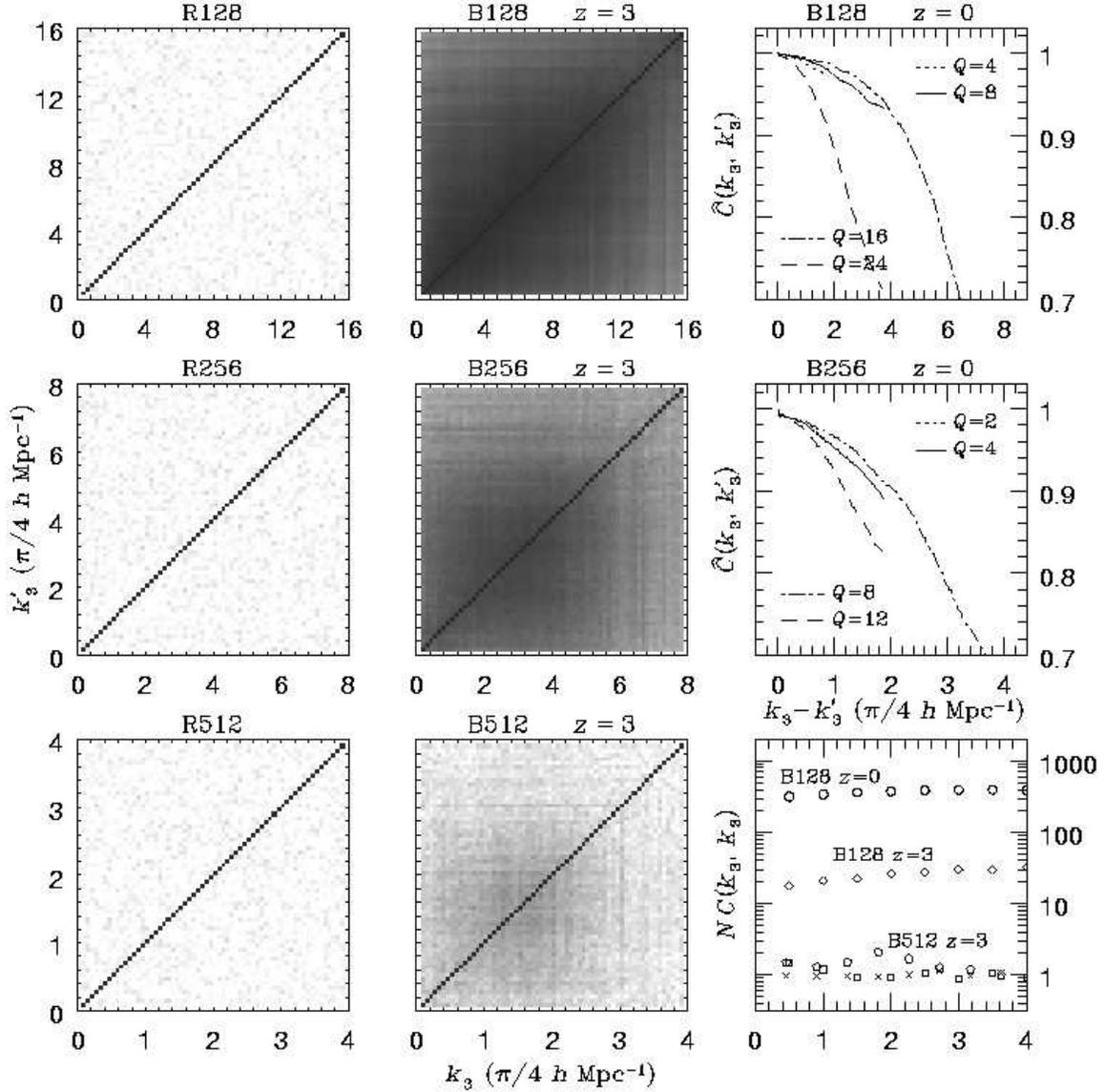}
\caption{The same as Fig.~\ref{fig:cov64}, except that the covariances
$\hat{C}(k_3,k'_3)$ and $C(k_3, k_3)$ are averaged over 256 groups, 
each of which consists of 1024 LOS's ($N=1024$).
\label{fig:cov1024}}
\end{figure}

\begin{figure} 
\centering
\includegraphics[width=6in]{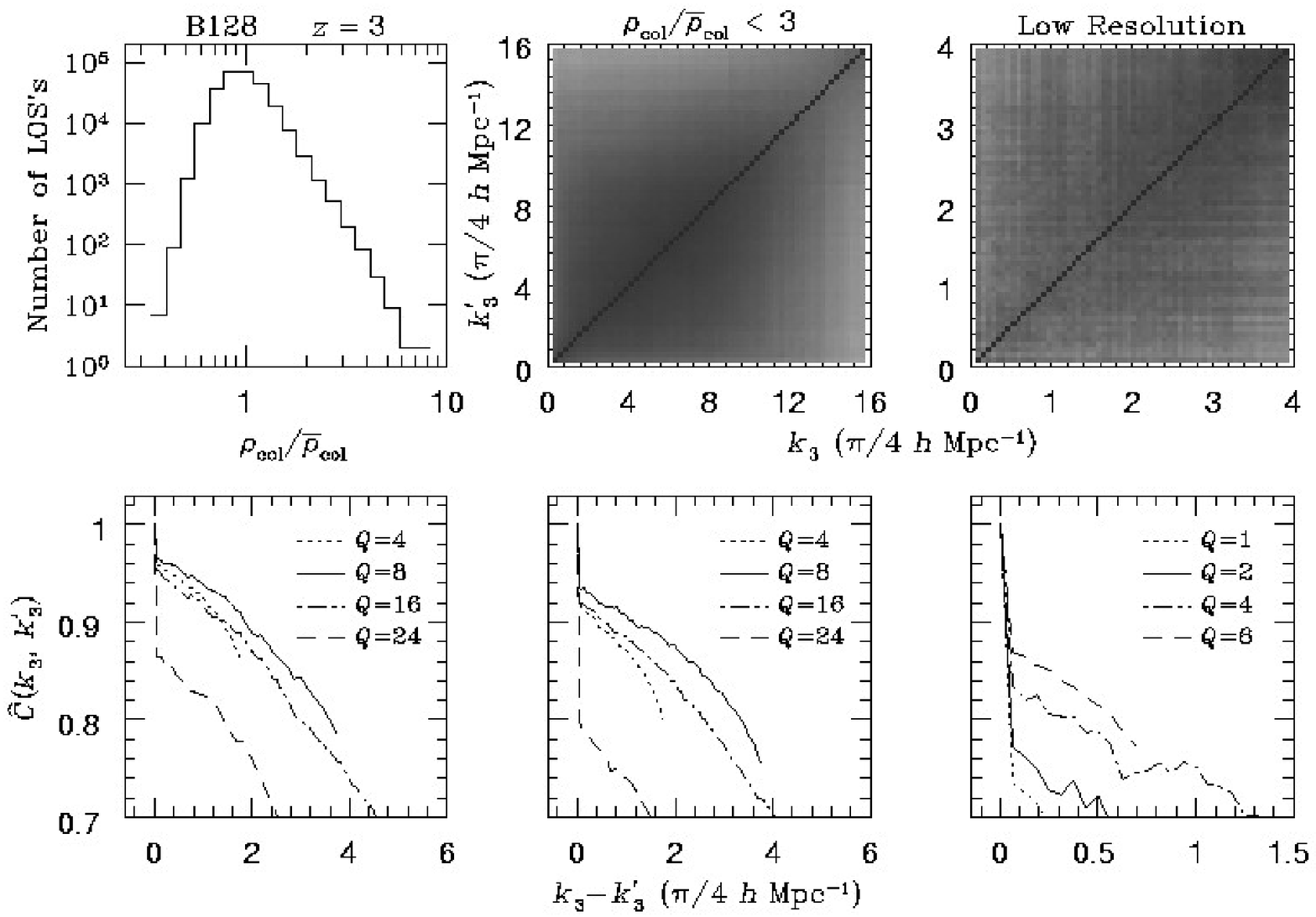}
\caption{The effect of the non-Gaussianity and resolution on the 
covariance. The left
column shows the distribution function of column density $\rho_{\rm
col}/\bar{\rho}_{\rm col}$ (upper panel) and cross sections of the
reduced covariance $\hat{C}(k_3, k_3')$ along $Q=(k_3+k_3')/(\pi/4\ h\
\mbox{Mpc}^{-1})$ (lower panel) for B128 at $z = 3$. The middle column
shows $\hat{C}(k_3, k_3')$ (upper panel) and its cross sections (lower
panel) for the same simulation output but with an exclusion of 318 LOS's 
that have $\rho_{\rm col}/\bar{\rho}_{\rm col} \ge 3$. The reduced
covariances in the first two columns are calculated in the same way as in
Fig.~\ref{fig:cov64}, i.e.~with 2000 groups and $N=64$. Thus, about half
of the 318 LOS's would be selected without the criterion $\rho_{\rm
col}/\bar{\rho}_{\rm col} < 3$. The right column is similar to the middle
column, but the density is assigned on a grid of $128^3$ nodes. All the
$128^2$ LOS's are selected and divided into 256 groups with $N=64$. The
grey scale is the same as Fig.~\ref{fig:cov64}.
\label{fig:cov64l}}
\end{figure}

Fig.~\ref{fig:var1d} shows the one-dimensional PS measured by averaging
over 64 and 1024 LOS's from the B128 simulation. Although the mean PS of
all the $512^2$ LOS's agrees with the result of a direct summation of the
three-dimensional PS using equation (\ref{eq:321fsum}), the deviation of
the PS for any particular group of 64 or 1024 LOS's is substantial,
especially at $z = 0$. The variance is smaller at $z = 3$ than at 
$z = 0$ because the cosmic density field is more Gaussian earlier on, 
and it is roughly 
inversely proportional to the number of LOS's. This can be seen better 
by comparing the lower right panel of Fig.~\ref{fig:cov64} with that of
Fig.~\ref{fig:cov1024}. However, even at $z = 3$ the variance of the 
one-dimensional PS is still much higher than the variance for GRFs, 
$N^{-1}P_{\rm 1D}^2(k_3)$, which indicates a heavy contribution from the 
trispectrum. The formulae in
Sections \ref{sec:321}--\ref{sec:short} often assume that LOS's are
sampled on a grid with fixed spacing, which may not be applicable to 
realistic data
such as inverted densities from the Ly$\alpha$ forest \citep{nh99,z03}.
Therefore, we sample the LOS's in two ways in Fig.~\ref{fig:var1d}: grid 
sampling and random sampling. Since no significant difference is observed, 
random sampling can be safely applied in the rest of this paper.

\begin{figure} 
\centering
\includegraphics[width=3in]{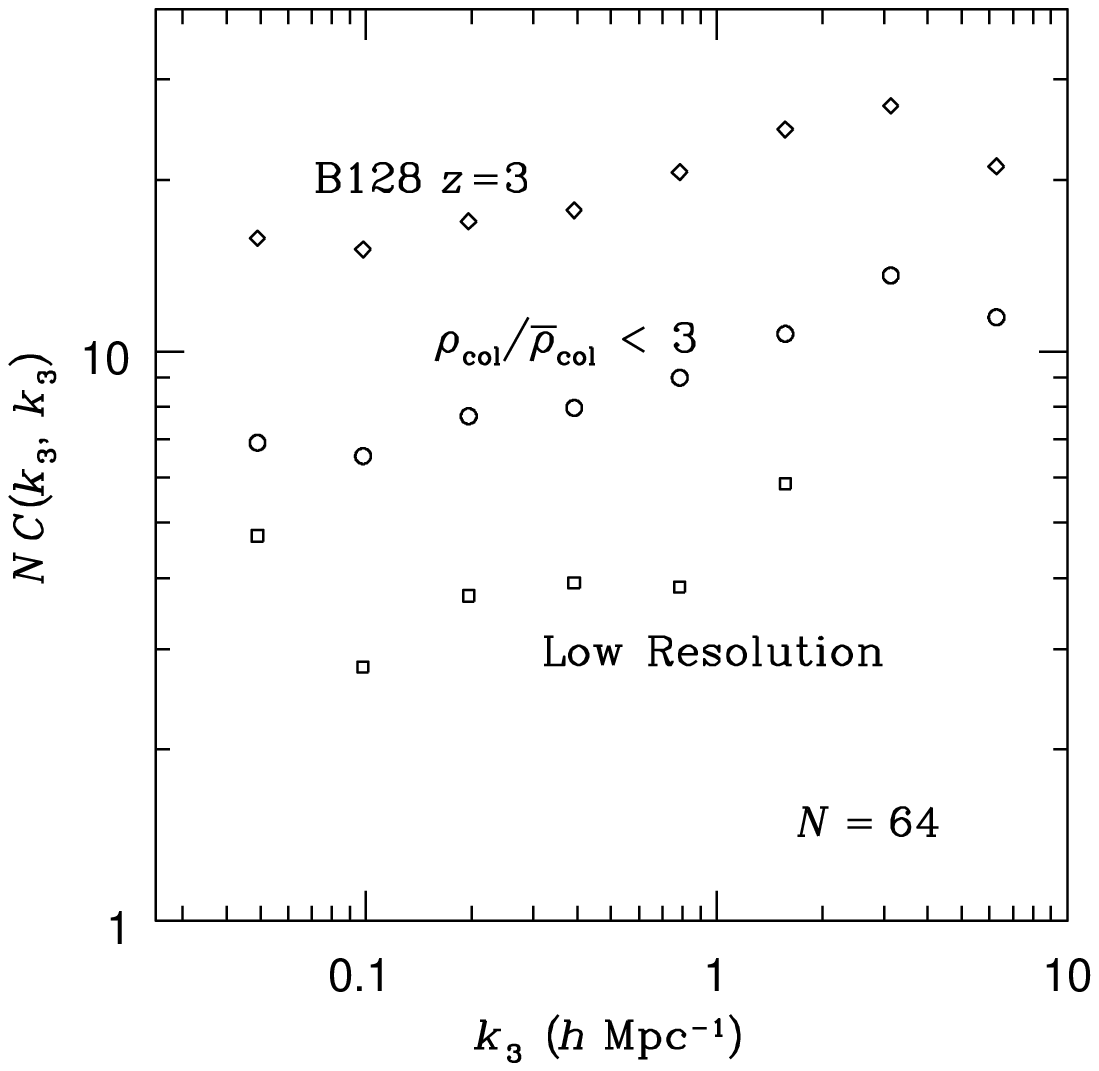}
\caption{The effect of the non-Gaussianity and resolution on the variance, 
i.e.~diagonal elements of the covariance $\sigma^2_{\rm 1D}(k_3, k'_3)$. 
Diamonds are the normalized variance $C(k_3, k_3)$ (multiplied by $N$) for 
B128 at $z=3$. Circles correspond to the middle column of 
Fig.~\ref{fig:cov64l}, which imposes the selection criterion 
$\rho_{\rm col} / \bar{\rho}_{\rm col} < 3$. Squares are from the 
low-resolution calculation in the right column of Fig.~\ref{fig:cov64l}.
\label{fig:var64l}}
\end{figure}

\begin{figure} 
\centering
\includegraphics[width=6in]{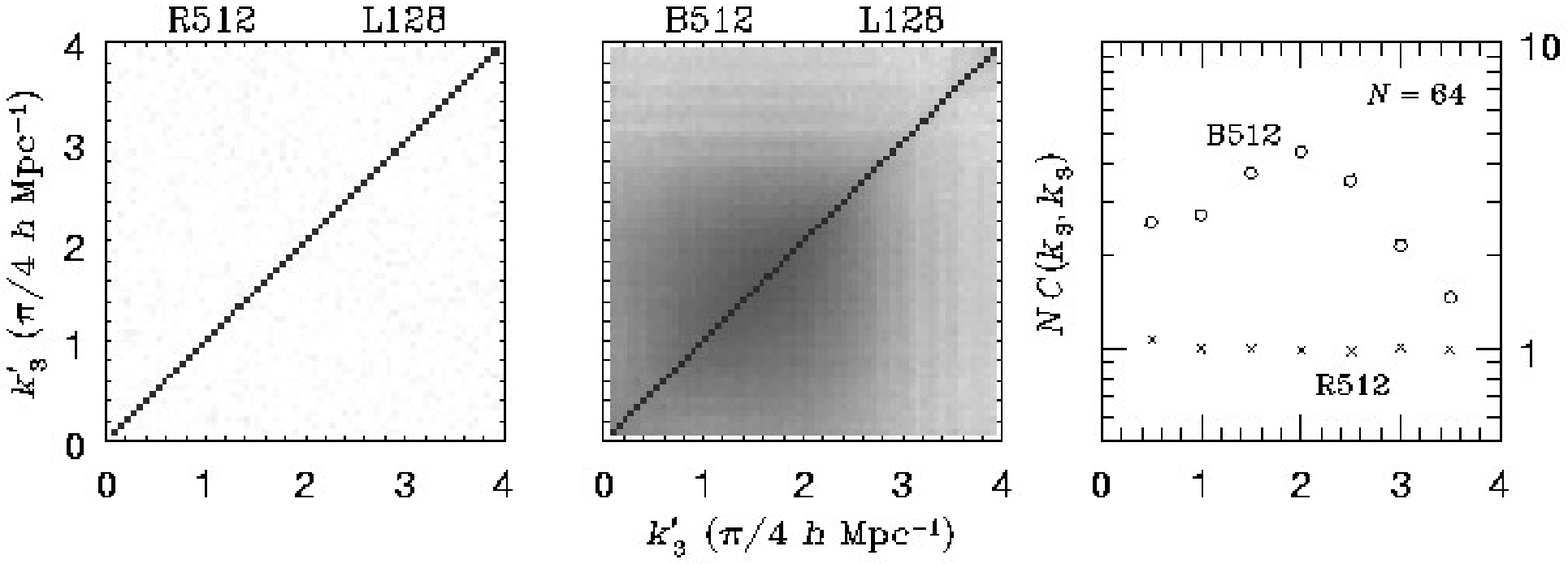}
\caption{The same as the bottom row of Fig.~\ref{fig:cov64}, but with 
length of LOS's $L = 128\ h^{-1}$Mpc. \label{fig:covlen}}
\end{figure}

The normalized covariances $\hat{C}(k_3, k_3')$ and $C(k_3,k_3')$ are
quantified in Figs.~\ref{fig:cov64} ($N=64$) and \ref{fig:cov1024} 
($N=1024$). The left column in each figure is the covariances of 
the spatially averaged one-dimensional PS from a single GRF that 
has the same box size and three-dimensional PS as
its corresponding simulation. Clearly, the covariances based on 
spatial average are nearly diagonal with unity diagonal elements. This is 
in agreement with the expectations based on ensemble average for GRFs and 
is consistent with the
ergodicity argument. The middle column is similar to the left column
except that the density fields are from simulations at $z = 3$. The modes
in the simulated density fields are strongly correlated, so that the
covariances are no longer diagonal. In other words, the trispectrum is
non-vanishing for the cosmic density field, as it is the only term that 
contributes to off-diagonal elements in the covariance.
The cosmic density field becomes so
non-Gaussian at $z=0$ that grey scale figures of the covariances will not
be readable. Hence, we only plot four
cross sections perpendicular the diagonal with 
$Q=(k_3+k_3')/(\pi/4\ h\ \mbox{Mpc}^{-1})$ for B128 
and B256 in the right column. The dominance of the diagonals suggested by 
these cross sections is actually weaker than that in the middle column, 
which can be seen by contrasting the cross sections for B128 at $z=0$ in 
Fig.~\ref{fig:cov64} with that for B128 at $z=3$ in Fig.~\ref{fig:cov64l}.

The non-Gaussianity is reflected not only in
the correlations between different modes but also in the
variance of the one-dimensional PS, i.e.~the diagonal elements of the 
normalized covariance $C(k_3, k'_3)$. The lower
right panels of Figs.~\ref{fig:cov64} and \ref{fig:cov1024} compare
$C(k_3, k_3)$ for five different density fields. The variance from 
the simulation B128 is orders of magnitude higher than the Gaussian value 
of $N^{-1}P^2_{\rm 1D}(k_3)$, and it grows
as the non-Gaussianity becomes stronger toward $z = 0$. As a result,
the sample variance error estimated for GRFs is much lower than what
can be actually measured from the cosmic density field. According to 
equation (\ref{eq:cov-tri}), both aliasing and the trispectrum contribute
to the variance. Since the GRFs 
have the same three-dimensional PS and are sampled in the same way as 
the simulations, their near-Gaussian variances of spatially averaged
one-dimensional PS's suggest that the contribution of the aliasing 
effect is negligible. Comparisons of
$C(k_3, k_3)$ for the same density field but with different sizes of 
sample ($N=64$ and 1024) confirm the observation in 
Fig.~\ref{fig:var1d} that the variance of the one-dimensional PS scales 
roughly as $N^{-1}$. This means that even
though the non-Gaussianity drives up the sample variance error of the
measured one-dimensional PS, one can still reduce the error by sampling a 
large number of LOS's.

The statistics of a large sample may sometimes be dominated by 
a small fraction of extreme cases. It is inevitable that some of the 
LOS's in our sample go through high-density halo structures, which may 
have significant contributions to the covariance matrices. To determine 
if a LOS falls in this category one has to assess the density 
profile intercepted by the LOS. We follow a much simpler approach by 
imposing a threshold to the mass column densities $\rho_{\rm col}$ of 
the LOS's. The one-point distribution function of column density contrast 
$\rho_{\rm col} / \bar{\rho}_{\rm col}$ from the simulation B128 
(upper left panel of Fig.~\ref{fig:cov64l}) peaks at 
$\rho_{\rm col} / \bar{\rho}_{\rm col} = 1$ as expected, and there 
clearly is a long-tail distribution (as compared to Gaussian or log-normal
distributions) of $\rho_{\rm col} / \bar{\rho}_{\rm col} \ga$ a few.
We re-calculate the covariance $\hat{C}(k_3, k_3')$ with a
selection criterion that the column density of each LOS $\rho_{\rm col} /
\bar{\rho}_{\rm col} < 3$. This criterion excludes 318 LOS's from 
the selection of LOS's for the statistics. As a total of 
$2000 \times 64$ LOS's are randomly selected from $512^2$ candidates, 
only half of the 318 LOS's, i.e. 0.1\% of the sample, would be selected
without the criterion. 
The result is shown in the upper middle panel
of Fig.~\ref{fig:cov64l}, which is visually similar to the 
B128 $z = 3$ panel in Fig.~\ref{fig:cov64}. 
However, the cross sections of $\hat{C}(k_3, k_3')$ with the 
selection criterion (lower middle panel of Fig.~\ref{fig:cov64l}) do
demonstrate a several-percent reduction of the off-diagonal 
elements in contrast to those without the selection criterion (lower 
left panel). In other words, the 0.1\% 
$\rho_{\rm col} / \bar{\rho}_{\rm col} \ge 3$ LOS's contribute at least
10 times as much to the covariance per LOS as 
$\rho_{\rm col} / \bar{\rho}_{\rm col} < 3$ LOS's do.
It is also useful to have a comparison of the normalized 
covariance $C(k_3, k'_3)$ because, unlike the reduced covariance 
$\hat{C}(k_3, k'_3)$ that only tells the relative strength between the 
diagonal and off-diagonal elements, $C(k_3, k'_3)$ reflects absolute
values of the covariance up to a factor of the product of PS's at $k_3$ 
and $k'_3$. Fig.~\ref{fig:var64l} shows the diagonal 
elements of the normalized covariance for B128. It is seen that the 
variance $C(k_3, k_3)$ of the one-dimensional PS is indeed reduced by
roughly a factor of 2 on all scales when the $0.1\%$
$\rho_{\rm col} / \bar{\rho}_{\rm col} \ge 3$ LOS's are excluded.

For a fixed number of particles, the simulation box sets a cut-off scale,
below which fluctuations cannot be properly represented in low-density 
regions such as the Ly$\alpha$ forest. In other words, the
number of particles and the size of the simulation determine the
highest-wavenumber modes that are included in calculations of the
one-dimensional PS and its covariance. Since the non-Gaussianity is
stronger at smaller scales, a larger simulation box cuts off more 
small-scale fluctuations and may cause the correlation
to appear weaker in Figs.~\ref{fig:cov64} and \ref{fig:cov1024}.
To test this, we assign the density field of the simulation B128 at $z=3$
on a grid of $128^3$ nodes. The spatial resolution is the same as the
simulation B512 on a grid of $512^3$ nodes. The covariance $\hat{C}(k_3,
k_3')$ is calculated in the same way as those in Fig.~\ref{fig:cov64} but
with fewer groups of LOS's. Each group still has 64 LOS's. The results are
shown in the right column of Fig.~\ref{fig:cov64l} and in 
Fig.~\ref{fig:var64l}. Evidently, there is a 
significant reduction of the correlations between different modes as well
as the variance of the one-dimensional PS. Indeed, the variance from the 
low-resolution calculation is close to that from the B512 simulation.
Thus, the apparent resemblance between B512 and the GRF R512 in 
Figs.~\ref{fig:cov64} and \ref{fig:cov1024} is mostly due to 
the low resolution of the large-box simulation.

It is expected from equation (\ref{eq:cov-short}) that the covariance
matrix will not be diagonal if the length of LOS's is less than the size
of the simulation box. Fig.~\ref{fig:covlen} shows the covariances that
are calculated in the same way as those in Fig.~\ref{fig:cov64} for the
GRF R512 and the simulation B512, except that each LOS is only 128
\mbox{$h^{-1}$Mpc} long. The effect of the length is not visible for the
GRF, but it does increase the correlation between different modes and 
doubles the variance of the one-dimensional PS for the simulation B512.
For real observations, the LOS length is always much less than the size 
of the observable universe, so that the window function 
$w(k_3,\tilde{k}_3)$ in the LOS direction will cause stronger mixing of 
modes and more pronounced increase of correlation and variance.

\section{Conclusions} \label{sec:con}

We have investigated the covariance of the one-dimensional mass PS for
both GRFs and simulated density fields. The non-linear evolution and the
non-Gaussianity of the cosmic density field on small scales \citep[see
also][]{zjf01, spj03, zsh03, z03} have caused the correlation between the
fluctuations on different scales and increased the cosmic variance of the
one-dimensional PS. Because of this, large number of LOS's are needed to
accurately measure the one-dimensional PS and recover the 
three-dimensional PS. 
% alias is a few times the PS but non-Gauss hundreds

The length of LOS's introduces a window function in the line-of-sight 
direction, 
which mixes neighboring Fourier modes in the cosmic density field. 
Fig.~\ref{fig:covlen} has demonstrated for simulations that the length 
of LOS's (a quarter of the simulation box size) does affect the covariance 
of the one-dimensional PS. 
The covariance of the observed one-dimensional PS will receive 
more contributions from this effect, because in practice the 
length of LOS's suitable for study is always much less than the size 
of the observable universe.

One may reduce the cosmic variance by binning \emph{independent}
Fourier modes. However, since the modes of the cosmic density field are
strongly correlated, binning will be less effective in reducing the sample
variance error. On the other hand, the non-Gaussian behaviour of the 
covariance provides important information of the field such as the 
trispectrum.

Our analysis shows that even with 64 LOS's at $z = 0$, the cosmic 
variance on the PS can be several times of the PS itself 
(see Fig.~\ref{fig:cov64}). 
Thus, the extra power on the scale of $128\ h^{-1}$Mpc ($\Omega = 1, 
\Omega_\Lambda = 0$) measured by \citet{bek90} from 4 samples of 
pencil-beam galaxy surveys can still be consistent with standard cosmogonies 
\citep[with considerations of effects such as redshift distortion and 
non-Gaussianity, see e.g.][]{kp91, pg91, vdw91, a94}.
One may also infer the one-dimensional density field and measure the PS
from the Ly$\alpha$ forest. Since the 
Ly$\alpha$ flux is the direct observable, many works have been based on 
flux statistics, which are then used along with simulations to constrain 
cosmology. Using $N$-body simulations and the 
pseudo-hydro technique, \citet{z03} find that the mapping between the 
nearly-Gaussian Ly$\alpha$ flux PS \citep{z04} and the underlying 
one-dimensional mass PS (in redshift space) has a much larger scatter than 
that between the flux PS and the one-dimensional
linear mass PS. This is expected because the one-dimensional PS 
of the cosmic density field has a variance much higher than 
the Gaussian variance. Therefore, one can measure the Ly$\alpha$ flux PS
precisely with a small sample of LOS's, but to recover the mass PS to
the same precision more LOS's are needed.

\section*{Acknowledgments}
We thank R.~Dav\'{e} for useful discussions and the referee for 
helpful comments. DJE was supported 
by NSF AST-0098577 and an Alfred P.~Sloan Research Fellowship.


\begin{thebibliography}{}

\bibitem[\protect\citeauthoryear{Amendola}{1994}]{a94} Amendola L.,
1994, ApJ, 430, L9

\bibitem[\protect\citeauthoryear{Bertschinger}{2001}]{b01} Bertschinger 
E., 2001, ApJS, 137, 1

\bibitem[\protect\citeauthoryear{Broadhurst et al.}{1990}]{bek90}
Broadhurst T.~J., Ellis R.~S., Koo D.~C., Szalay A.~S., 1990, 
Nature, 343, 726

\bibitem[\protect\citeauthoryear{Cooray \& Hu}{2001}]{ch01} Cooray A., 
Hu W., 2001, ApJ, 554, 56

\bibitem[\protect\citeauthoryear{Croft et al.}{1998}]{cwk98} 
Croft R.~A.~C., Weinberg D.~H., Katz N., Hernquist L., 1998, ApJ, 495, 44

\bibitem[\protect\citeauthoryear{Croft et al.}{1999}]{cwp99} 
Croft R.~A.~C., Weinberg D.~H., Pettini M., Hernquist L., Katz N., 1999, 
ApJ, 520, 1

\bibitem[\protect\citeauthoryear{Croft et al.}{2002}]{cwb02} 
Croft R.~A.~C., Weinberg D.~H., Bolte M., Burles S., Hernquist L., 
Katz N., Kirkman D., Tytler D., 2002, ApJ, 581, 20

\bibitem[\protect\citeauthoryear{Feldman, Kaiser \& Peacock}{1994}]
{fkp94} Feldman H.~A., Kaiser N., Peacock J.~A., 1994, ApJ, 426, 23

\bibitem[\protect\citeauthoryear{Gnedin \& Hamilton}{2002}]{gh02}
Gnedin N.~Y., Hamilton A.~J.~S., 2002, MNRAS, 334, 107

\bibitem[\protect\citeauthoryear{Hockney \& Eastwood}{1981}]{he81} 
Hockney R.~W., Eastwood J.~W., 1981, Computer Simulation Using Particles,
McGraw-Hill, New York, NY, p.~152

\bibitem[\protect\citeauthoryear{Hui, Stebbins \& Burles}{1999}]{hsb99}
Hui L., Stebbins A., Burles S., 1999, ApJ, 511, L5

\bibitem[\protect\citeauthoryear{Kaiser \& Peacock}{1991}]{kp91}
Kasier N., Peacock J.~A., 1991, ApJ, 379, 482

\bibitem[\protect\citeauthoryear{Kim et al.}{2004}]{kvh04} Kim T.-S., 
Viel M., Haehnelt M.~G., Carswell R.~F., Cristiani S., 2004, MNRAS, 
347, 355

\bibitem[\protect\citeauthoryear{Lumsden, Heavens \& Peacock}{1989}]
{lhp89} Lumsden S.~L., Heavens A.~F., Peacock J.~A., 1989, MNRAS, 238, 
293

\bibitem[\protect\citeauthoryear{Ma \& Bertschinger}{1995}]{mb95} 
Ma C.-P., Bertschinger E., 1995, ApJ, 455, 7

\bibitem[\protect\citeauthoryear{Mandelbaum et al.}{2003}]{mms03} 
Mandelbaum R., McDonald P., Seljak U., Cen R., 2003, MNRAS, 344, 776

\bibitem[\protect\citeauthoryear{McDonald}{2003}]{m03} McDonald P.,
2003, ApJ, 585, 34

\bibitem[\protect\citeauthoryear{McDonald \& Miralda-Escud\'{e}}{1999}]
{mm99} McDonald P., Miralda-Escud\'{e} J., 1999, ApJ, 518, 24

\bibitem[\protect\citeauthoryear{McDonald et al.}{2004}]{msb04}
McDonald P. et al., 2004, submitted to ApJ, astro-ph/0405013

\bibitem[\protect\citeauthoryear{Meiksin \& White}{1999}]{mw99} 
Meiksin A., White M., 1999, MNRAS, 308, 1179

\bibitem[\protect\citeauthoryear{Nusser \& Haehnelt}{1999}]{nh99} 
Nusser A., Haehnelt M., 1999, MNRAS, 303, 179

\bibitem[\protect\citeauthoryear{Park \& Gott}{1991}]{pg91}
Park C., Gott, J.~R. III, 1991, MNRAS, 249, 288

\bibitem[\protect\citeauthoryear{Peebles}{1980}]{p80} Peebles P.~J.~E.,
1980, The Large Scale Structure of the Universe, Princeton Univ. Press, 
Princeton, NJ, p.~150

\bibitem[\protect\citeauthoryear{Rollinde et al.}{2003}]{rpp03} 
Rollinde E., Petitjean P., Pichon C., Colombi S., Aracil B., 
D'Odorico V., Haehnelt M.~G., 2003, MNRAS, 341, 1279

%\bibitem[\protect\citeauthoryear{Shannon}{1949}]{s49} Shannon C.~E., 
%1949, Proc. IRE, 37, 10

\bibitem[\protect\citeauthoryear{Smith et al.}{2003}]{spj03}
Smith R.~E. et al., 2003, MNRAS, 341, 1311

\bibitem[\protect\citeauthoryear{Spergel et al.}{2003}]{svp03} 
Spergel D.~N. et al., 2003, ApJS, 148, 175

\bibitem[\protect\citeauthoryear{Springel, Yoshida \& White}{2001}]
{syw01} Springel V., Yoshida N., White S.~D.~M., 2001, New Astr., 6, 79

%\bibitem[\protect\citeauthoryear{Unser}{2000}]{u00} Unser M., 2000, 
%Proc. IEEE, 88, 569

\bibitem[\protect\citeauthoryear{van de Weygaert}{1991}]{vdw91}
van de Weygaert R., 1991, MNRAS, 249, 159

\bibitem[\protect\citeauthoryear{Viel et al.}{2002}]{vmm02} Viel M., 
Matarrese S., Mo H.~J., Haehnelt M.~G., Theuns T., 2002, MNRAS, 329, 848

\bibitem[\protect\citeauthoryear{Zaldarriaga, Scoccimarro \& Hui}{2003}]
{zsh03} Zaldarriaga M., Scoccimarro R., Hui L., 2003, ApJ, 590, 1

\bibitem[\protect\citeauthoryear{Zhan}{2003}]{z03} Zhan H., 2003, MNRAS, 
344, 935

\bibitem[\protect\citeauthoryear{Zhan, Jamkhedkar \& Fang}{2001}]{zjf01} 
Zhan H., Jamkhedkar P., Fang L.-Z., 2001, ApJ, 555, 58

\bibitem[Zhan(2004)]{z04} Zhan, H. 2004, PhD Thesis, 
Univ.~of Arizona, astro-ph/0408379
\end{thebibliography}
\end{document}